\documentclass[aps,pre,twocolumn,groupedaddress,showpacs]{revtex4}
\usepackage{amsfonts}
\usepackage{graphicx}
\begin{document}
\newcommand{\ud}{{\mathrm d}}
\newcommand{\sech}{\mathrm{sech}}

\title{Flux reversal in a simple random walk model on a fluctuating
symmetric lattice}


\author{Jes\'us Casado-Pascual}
\affiliation{F\'{\i}sica Te\'orica,
Universidad de Sevilla, Apartado de Correos 1065, Sevilla 41080, Spain}

\date{\today}

\begin{abstract}
A rather simple random walk model on a one-dimensional lattice is put
forward. The lattice as a whole switches randomly between two possible
states which are spatially symmetric. Both lattice states are
identical, but translated by one site with respect to each other, and
consist of infinite arrays of absorbing sites separated by two
non-absorbing sites. Exact explicit expressions for the long-time
velocity and the effective diffusion coefficient are obtained and
discussed. In particular, it is shown that the direction of the steady
motion can be reversed by conveniently varying the values of either
the mean residence times in the lattice states or the transition rates
to the absorbing and non-absorbing sites.

\end{abstract}

\pacs{05.40.Fb, 05.60.Cd, 87.16.Nn}

\maketitle
\section{Introduction}
The study of the transport of Brownian particles along periodic
structures has attracted considerable and ever growing interest in the
last years due to its relevance in a wide variety of problems in
physics, chemistry, biology, and nanotechnology \cite{Reimann,
Hanggi1}.  As required by the second law of thermodynamics, Brownian
motion cannot induce a steady directed motion in a system at
equilibrium (see, for instance, Ref.~\cite{Reimann} for a detailed
proof). Nevertheless, a Brownian particle can display a directed
motion in a periodic potential which switches randomly or periodically
between two or more states
\cite{Bier,Prost,Chauwin,Rousselet,Faucheux}. For this to be possible,
certain spatiotemporal symmetry and supersymmetry conditions must be
broken \cite{Reimann1}. An important phenomenon 
often observed in these systems is the reversal of the current direction 
upon variation of a system parameter \cite{Doering,Elston}. Though in many cases the spatiotemporal symmetry breaking is achieved by using a spatially asymmetric periodic
potential, in Ref.~\cite{Chen} it was shown that it can also be
obtained by using a spatially symmetric periodic potential which
fluctuates between three states, one of them being of free diffusion.
As shown in Ref.~\cite{Kanada}, if the symmetry of glide
 reflection is broken, a directed transport can also be induced in a spatially 
symmetric potential which switches randomly between two states.  Besides the purely theoretical interest, the study of this kind of
models is relevant because it may provide valuable information about
the workings of the so-called molecular motors
\cite{Reimann,Howard}. Molecular motors are able to travel along
polymer filaments by utilizing chemical energy from ATP hydrolysis,
and are responsible for the transport of different substances inside
the eukaryotic cells.

Recently, it has been demonstrated experimentally that the Brownian
motion of a colloidal sphere can be rectified using a periodic
potential that alternates between three \cite{Grier1} or two
\cite{Grier2} states which differ only by a discrete translation. Each
state is spatially symmetric and consists of a large array of optical
traps, created with dynamic holographic optical tweezers, each of which can localize a colloidal sphere. The pattern of discrete optical traps can be modeled as a periodic array of Gaussian wells of width $\sigma$ and depth $V_0$, uniformly separated by a distance $L$ substantially larger than $\sigma$. 
In the three-state case considered in Ref.~\cite{Grier1}, the initial array of optical traps is extinguished after time $T$ and replaced immediately with a second array, which is displaced from the first by $L/3$. This process is repeated after time $T$ with an additional displacement of $L/3$. A cycle is completed when, after time $T$, the array of traps is displaced once again by $L/3$ and returns to its initial state. In the two-state case reported in Ref.~\cite{Grier2}, the first step is the same as in \cite{Grier1}, i.e., the initial array of optical traps is extinguished after time $T_1$ and replaced immediately with a second array, which is displaced from the first by $L/3$. This second array is extinguished after time $T_2$ and replaced again by the first, thereby completing one cycle. One main
result of the three-state case is that the direction of the motion
depends sensitively on the duration of the states, $T$, relative to the time
required for the particle to diffuse the intertrap separation
\cite{Grier1}. By contrast, in the two-state case this direction
depends on the relative durations of the two states $T_2/T_1$
\cite{Grier2}. Specifically, the flux is directed from each optical trap in the longer-duration state toward the
nearest optical trap in the short-lived state, and vanishes for $T_1=T_2$. Thus, in both cases, by varying the value of the
corresponding parameter the flux can be reversed. 

To gain a deeper
insight into the behavior of these systems, it would be desirable to
develop simplified models, amenable to analytical treatment, which
capture the key features of the original complex systems. The main focus of this paper
 is to present such a simplified model for the experiment described in Ref.~\cite{Grier2}.
The paper is organized as follows. In the following section, we introduce the model and provide definitions of the quantities of interest, namely, the long-time velocity and the effective diffusion coefficient. In Sec.~\ref{sect3}, we obtain explicit expressions for these two quantities. The main results arising from these expressions are discussed in Sec.~\ref{sect4}. Finally, we present conclusions for the main findings of our work.

\section{Description of the model}

In our simplified model the motion of the colloidal sphere in the periodic potential is
modelled by a random walk on a one-dimensional lattice with its sites
located at $x_n=n h$, with $n\in \mathbb{Z}$, where it has been
assumed that the distance between two consecutive sites
$x_{n+1}-x_n=h$ is independent of $n$. The lattice can be in two
possible states that will be indicated by the label $\alpha=\pm 1$. In
contrast with the approach described in Ref.~\cite{Grier2} in which
the potential switches periodically between its two states, here we
will assume that the lattice fluctuates randomly between its two
possible states. The evolution of the joint probabilities
$p_n^{(\alpha)}(t)$ that the particle is at site $x_n$ and the lattice
in state $\alpha$ at time $t$ are governed by the master equation
\begin{eqnarray}
\label{equ2}
\dot{p}_n^{(\alpha)}(t)&=&-\left[
L_{n}^{(\alpha)}+R_{n}^{(\alpha)}\right]
p_n^{(\alpha)}(t)+R_{n-1}^{(\alpha)} p_{n-1}^{(\alpha)} (t)\nonumber
\\&&+L_{n+1}^{(\alpha)} p_{n+1}^{(\alpha)}(t)+\gamma^{(-\alpha)}
p_n^{(-\alpha)}(t)-\gamma^{(\alpha)}
p_n^{(\alpha)}(t),\nonumber\\
&&
\end{eqnarray}  
where $\gamma^{(\alpha)}$ is the transition rate from the lattice
state $\alpha$ to $-\alpha$, and $L_{n}^{(\alpha)}$ and $R_{n}^{(\alpha)}$ 
are the transition rates from site $x_n$ to site $x_{n-1}$ and $x_{n+1}$,
respectively, assuming that the lattice is in state $\alpha$. The
initial condition is irrelevant since we are interested in the
long-time behavior of the system.  

\begin{figure}
\includegraphics[width=8cm]{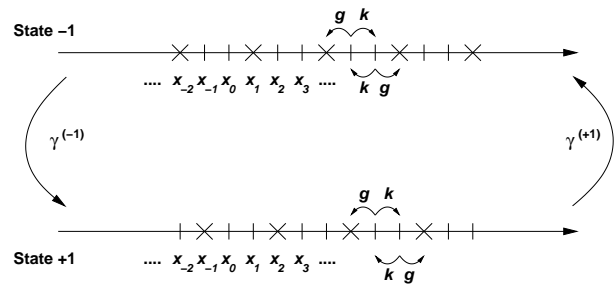}
\caption{\label{Fig1} Sketch of the fluctuating lattice. The absorbing
sites are depicted by crosses.}
\end{figure}

Retaining the main features of the model described in
Ref.~\cite{Grier2}, we will assume that both states of the lattice are
spatially symmetric (i.e., there exists a $n'\in \mathbb{Z}$ such that
$L_{n'+n}^{(\alpha)}=R_{n'-n}^{(\alpha)}$ for all $n\in \mathbb{Z}$)
and periodic with period $N$ (i.e.,
$L_{n+N}^{(\alpha)}=L_{n}^{(\alpha)}$ and
$R_{n+N}^{(\alpha)}=R_{n}^{(\alpha)}$ for all $n\in \mathbb{Z}$). This
is an important distinction between our model and other similar models
previously reported in the literature
\cite{Freund,Kolomeisky,Ambaye,Makhnovskii}, in which at least one of
the lattice states is spatially asymmetric.  For simplicity, we will
restrict our study to the case $N=3$, which is the smallest value of
the period for which a non-zero steady directed motion is possible
under the above conditions of spatial symmetry and periodicity
\cite{prove}.  The lattice state $\alpha=+1$ is obtained from the
lattice state $\alpha=-1$ by translating it to the right one site, so
that, $L^{(+1)}_{n}=L^{(-1)}_{n-1}$ and $R^{(+1)}_{n}=R^{(-1)}_{n-1}$.
When the lattice is in state $\alpha=-1$, there are absorbing sites at
the locations $x_{3 j+1}$, for all $j\in \mathbb{Z}$, and
consequently, $L^{(-1)}_{3 j+1}=R^{(-1)}_{3
j+1}=L^{(+1)}_{3j+2}=R^{(+1)}_{3 j+2}=0$.  These absorbing sites play
the role of the optical traps in our simplified model. The transition
rates to an absorbing and a non-absorbing site will be denoted by $g$
and $k$, respectively, so that, $L^{(-1)}_{3 j+2}=R^{(-1)}_{3
j}=L^{(+1)}_{3 j}=R^{(+1)}_{3 j+1}=g$ and $R^{(-1)}_{3
j+2}=L^{(-1)}_{3 j}=R^{(+1)}_{3j}=L^{(+1)}_{3j+1}=k$. A sketch of the
model just described is shown in Fig.~\ref{Fig1}.

\begin{figure}
\includegraphics[width=8cm]{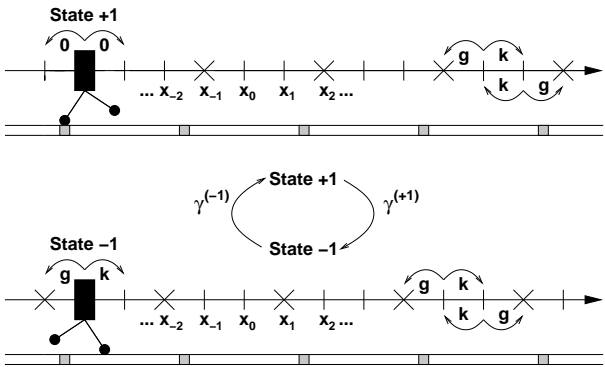}
\caption{\label{Fig2} An alternative interpretation of the model sketched 
in Fig.~\ref{Fig1}. The binding and absorbing
sites are depicted by grey squares and crosses, respectively.}
\end{figure}

It is worthwhile to point out that the model just described is
formally equivalent to another random walk model which mimics the motion of 
some molecular motors. In this model the walker switches
 randomly between two possible states, indicated by the label 
$\alpha=\pm 1$, $\gamma^{(\alpha)}$ being the transition rate from 
the state $\alpha$ to $-\alpha$ (see Fig.~\ref{Fig2}). These states are 
characterized by the position of two legs, each with a head-shaped object 
attached at one of their ends. The walker ``walks'' on a rod which possesses an 
infinite array of binding sites, depicted by grey squares in Fig.~\ref{Fig2}. 
These binding sites are separated from each other by a distance $3 h$, and 
translated by $h/2$ with respect to the sites of the lattice. Whenever any 
of the walker's ``heads'' comes in contact with a binding site, the walker stays
trapped at that site until a fluctuation of its state occurs. For each
state of the walker, the sites of the lattice with this property
(absorbing sites) are depicted by crosses in Fig.~\ref{Fig2}. When
none of the walker's ``heads'' are in contact with a binding site, the
walker can hop to its nearest-neighbor sites. Then, the transition
rate from a site to its neighbors is $g$, if after hopping any of the
``heads'' comes in contact with a binding site, or $k$, otherwise. 
This interpretation of the model differs from
the one considered at the end of Ref.~\cite{Grier2}, in
three important aspects. First, ours is discrete and exactly
solvable. Second, in ours the walker fluctuates randomly between its two possible states, whereas in theirs it switches periodically between them.
 In this sense, our model may be more effective
for describing real molecular motors, since in this class of molecules
the directed motion arises as a consequence of a competition between
two stochastic time scales (the ATP binding rate and the ATP
hydrolysis rate). Third, in \cite{Grier2} the direction of the steady
motion depends only on the relative durations of the two states,
whereas here, as we will see later on, it also depends on the sign of $g-k$.

We are interested in the study of the long-time velocity or flux $v$
and the effective diffusion coefficient $D_{\mathrm{eff}}$, defined as
\begin{equation}
\label{vdef}
v=\lim_{t\rightarrow \infty} \frac{d}{d t}\langle x(t) \rangle
\end{equation}
and
\begin{equation}
\label{Ddef}
D_{\mathrm{eff}}=\frac{1}{2}\lim_{t\rightarrow \infty} \frac{d}{d
  t}\left[\langle x^2(t) \rangle -\langle x(t) \rangle^2 \right],
\end{equation}
where the $l$th moment of the position is given by $\langle x^l(t)
\rangle=h^l \sum_{\alpha=\pm 1} \sum_{n=-\infty}^{+\infty} n^l
\,p_n^{(\alpha)}(t)$.

There are certain properties of $v$ and $D_{\mathrm{eff}}$ that can be
predicted by simple considerations. For instance, in the absence of
fluctuations between the two lattice states, both the long-time
velocity and the effective diffusion coefficient will be zero, as the
particle ends up being trapped by an absorbing site. Even if the sites
depicted by crosses in Fig.~\ref{Fig1} were not absorbing and,
therefore, there were a non-zero transition rate from them (the same
to the left and to the right), the long-time velocity would be also
zero in the absence of fluctuations between the two lattice
states. This is so because, in that case, both lattice states would be
invariant under reflections about any of the sites with
crosses. Another interesting property is obtained if one takes into
account that the fluctuating periodic lattice sketched in
Fig.~\ref{Fig1} is invariant under the transformation
$\{n,\alpha\}\Rightarrow \{-n,-\alpha\}$. Consequently, the long-time
velocity and the effective diffusion coefficient fulfill the symmetry
relations
$v\big[\gamma^{(+1)},\gamma^{(-1)}\big]=-v\left[\gamma^{(-1)},
\gamma^{(+1)}\right]$ and
$D_{\mathrm{eff}}\left[\gamma^{(+1)},\gamma^{(-1)}\right]=D_{\mathrm{eff}}
\left[\gamma^{(-1)}, \gamma^{(+1)}\right]$, where we have written
explicitly the dependence of $v$ and $D_{\mathrm{eff}}$ on
$\gamma^{(+1)}$ and $\gamma^{(-1)}$. Thus, if
$\gamma^{(+1)}=\gamma^{(-1)}$, the long-time velocity will be zero.

\section{Explicit expressions for the long-time
velocity and the effective diffusion coefficient}

\label{sect3}

In order to go further in the analysis $v$ and $D_{\mathrm{eff}}$, it
is necessary to obtain explicit expressions for them. To do that, we
will extend the approach developed in Ref.~\cite{Derrida} to the case
of a fluctuating lattice. The procedure described here is quite
general and can be applied to any fluctuating periodic lattice with an
arbitrary period $N$. For this reason, in what follows it will be
described in general terms.  An alternative approach based on the
gradient expansion of the probability distribution corresponding to a
coarse-grained description of the random walk model can be found in
Ref.~\cite{Freund}. First, let us introduce the quantities
\begin{equation}
\label{qdef}
q_n^{(\alpha)}(t)=\sum_{l=-\infty}^{\infty}p_{n+l N}^{(\alpha)}(t)
\end{equation} 
and
\begin{equation}
\label{gdef}
s_n^{(\alpha)}(t)=h\sum_{l=-\infty}^{\infty}(n+l N)p_{n+l
N}^{(\alpha)}(t)-\langle x(t) \rangle q_n^{(\alpha)}(t),
\end{equation} 
which are periodic in $n$ with period $N$, i.e.,
$q_{n+N}^{(\alpha)}(t)=q_{n}^{(\alpha)}(t)$ and
$s_{n+N}^{(\alpha)}(t)=s_n^{(\alpha)}(t)$ for all $n\in
\mathbb{Z}$. Substituting the expressions for the first and second
moments of the position into Eqs.~(\ref{vdef}) and (\ref{Ddef}), and
replacing $\dot{p}_n^{(\alpha)}(t)$ by the right-hand side of
Eq.~(\ref{equ2}), it is not difficult to prove that the long-time
velocity and the effective diffusion coefficient can be expressed as
\begin{equation}
\label{vtdef}
v=h \sum_{\alpha=\pm 1} \sum_{n=1}^{N}
\left[R_n^{(\alpha)}-L_n^{(\alpha)}\right]Q_n^{(\alpha)}
\end{equation}
and
\begin{eqnarray}
\label{dtdef}
D_{\mathrm{eff}}&=&\frac{h^2}{2} \sum_{\alpha=\pm 1} \sum_{n=1}^{N}
\left[R_n^{(\alpha)}+L_n^{(\alpha)}\right]Q_n^{(\alpha)}\nonumber \\
&&+ h \sum_{\alpha=\pm 1} \sum_{n=1}^{N}
\left[R_n^{(\alpha)}-L_n^{(\alpha)}\right]S_n^{(\alpha)},
\end{eqnarray}
where $Q_n^{(\alpha)}=\lim_{t\rightarrow \infty}q_n^{(\alpha)}(t)$ and
$S_n^{(\alpha)}=\lim_{t\rightarrow \infty}s_n^{(\alpha)}(t)$. Thus,
for the evaluation of $v$ and $D_{\mathrm{eff}}$ it suffices to calculate
the values of the $4N$ numbers $Q_n^{(\alpha)}$ and $S_n^{(\alpha)}$,
with $n=1,\dots,N$ and $\alpha=\pm 1$.

Differentiating Eqs.~(\ref{qdef}) and (\ref{gdef}) with respect to
time, making use of Eq.~(\ref{equ2}), as well as of the periodicity of
$L_n^{(\alpha)}$ and $R_n^{(\alpha)}$, and taking into account that
$\lim_{t\rightarrow \infty}{\dot q}_n^{(\alpha)}(t)=\lim_{t\rightarrow
\infty}{\dot s}_n^{(\alpha)}(t)=0$, it is easy to see that the
long-time limits $Q_n^{(\alpha)}$ and $S_n^{(\alpha)}$ fulfill the set
of $4N$ algebraic equations
\begin{eqnarray}
\label{calcQ}
-\left[L_n^{(\alpha)}+R_n^{(\alpha)}\right]Q_n^{(\alpha)}\;+\;
R_{n-1}^{(\alpha)}Q_{n-1}^{(\alpha)}&+&L_{n+1}^{(\alpha)}
Q_{n+1}^{(\alpha)}\nonumber \\ +\;\gamma^{(-\alpha)}Q_n^{(-\alpha)}\;-\;
\gamma^{(\alpha)}Q_n^{(\alpha)}&=&0,\\
\label{calcS}
-\left[L_n^{(\alpha)}+R_n^{(\alpha)}\right]S_n^{(\alpha)}\;+\;
R_{n-1}^{(\alpha)}S_{n-1}^{(\alpha)}&+&L_{n+1}^{(\alpha)}
S_{n+1}^{(\alpha)}\nonumber \\ +\;\gamma^{(-\alpha)}S_n^{(-\alpha)}\;-\;
\gamma^{(\alpha)}S_n^{(\alpha)}&=&F_n^{(\alpha)},
\end{eqnarray}
for $n=1,\dots,N$ and $\alpha=\pm 1$, where
\begin{equation}
F_n^{(\alpha)}=h\left[L_{n+1}^{(\alpha)}Q_{n+1}^{(\alpha)}
  -R_{n-1}^{(\alpha)}Q_{n-1}^{(\alpha)}\right]+v\,Q_{n}^{(\alpha)},
\end{equation}
and where the periodic boundary conditions
$Q_0^{(\alpha)}=Q_N^{(\alpha)}$, $Q_{N+1}^{(\alpha)}=Q_1^{(\alpha)}$,
$S_0^{(\alpha)}=S_N^{(\alpha)}$, and
$S_{N+1}^{(\alpha)}=S_1^{(\alpha)}$ are assumed.

Summing up the $2 N$ equations contained either in the set of
equations (\ref{calcQ}) or (\ref{calcS}), it is easy to realize that
in each set there are only $2N-1$ independent equations (excluding
some pathological cases where the number can be even lower). Thus, to
univocally determine the $4 N$ quantities $Q_n^{(\alpha)}$ and
$S_n^{(\alpha)}$, it is also necessary to add the two additional
equations $\sum_{\alpha=\pm 1} \sum_{n=1}^{N} Q_n^{(\alpha)}=1$ and
$\sum_{\alpha=\pm 1} \sum_{n=1}^{N} S_n^{(\alpha)}=0$, which can be
straightforwardly proved from the definitions of $Q_n^{(\alpha)}$ and
$S_n^{(\alpha)}$.

To apply the above procedure to the model sketched in Fig.~\ref{Fig1},
first one has to solve this set of equations, with $N=3$ and the
corresponding values for the transition rates $L_n^{(\alpha)}$ and
$R_n^{(\alpha)}$. Replacing the results obtained for $Q_n^{(\alpha)}$
and $S_n^{(\alpha)}$ in Eqs.~(\ref{vtdef}) and (\ref{dtdef}), one
obtains after some lengthy simplifications that
\begin{equation}
\label{velocity}
v=3 h \Delta \rho \left(1-\Delta \rho^2\right)(g-k)g k \Gamma^2 A^{-1}
\end{equation}
and 
\begin{eqnarray}
\label{effdiff}
D_{\mathrm{eff}}&=&\frac{9}{2} h^2 \left(1-\Delta \rho^2\right)g k
(g+k)(2g+\Gamma)\Gamma A^{-1}\nonumber\\ &&-v^2 B \Gamma^{-1}A^{-1},
\end{eqnarray}
where
\begin{eqnarray}
\label{Adef}
A&=&4 g^2 (g+2k)^2+8 g (g+k)(g+2k)\Gamma\nonumber \\
&&+\left[4(g+k)^2-(1+3 \Delta \rho^2) k^2 \right]\Gamma^2,
\end{eqnarray}
\begin{eqnarray}
B&=&4 g^2 (g+2k)^2+24g (g+k)(g+2k)\Gamma\nonumber
\\&&+\left[28(g+k)^2-3(3+ \Delta \rho^2) k^2 \right]\Gamma^2+8(g+k)
\Gamma^3,\nonumber\\&&
\end{eqnarray}
$\Gamma=\gamma^{(+1)}+\gamma^{(-1)}$, and $\Delta
\rho=\rho_{\mathrm{eq}}^{(-1)}-\rho_{\mathrm{eq}}^{(+1)}$,
$\rho_{\mathrm{eq}}^{(\alpha)}$ being the equilibrium population of
the lattice state $\alpha$, i.e.,
$\rho_{\mathrm{eq}}^{(\alpha)}=\gamma^{(-\alpha)}/\Gamma$.  This
quantity can alternatively be expressed as $\Delta
\rho=\left[T^{(-1)}-T^{(+1)}\right]/\left[T^{(-1)}+T^{(+1)} \right]$,
where $T^{(\alpha)}=1/\gamma^{(\alpha)}$ is the mean residence time of
the lattice in the state $\alpha$.  

\section{Results}

\label{sect4}

An important result that follows from Eq.~(\ref{velocity}) is that the
direction of the steady motion not only depends on the sign of $\Delta
\rho$, as would be expected from the previous symmetry considerations,
but also on the sign of the difference between the transition rates to
an absorbing and a non-absorbing site, $g-k$. Thus, for instance, if
$T^{(-1)}>T^{(+1)}$, the steady directed motion is from left to right
for $g>k$, whereas it is from right to left for $g<k$. Obviously, if
$T^{(-1)}<T^{(+1)}$, the behavior is just the opposite. Finally, no
steady directed motion is possible if either $T^{(-1)}=T^{(+1)}$ or
$g=k$.  This subtle dependence of the direction of the steady motion
on the values of $g$ and $k$ cannot be inferred {\em a priori} from
symmetry considerations, and is one of the main results of this work.
It is important to point out that this model predicts correctly the
direction of the steady motion of the experiment described in
Ref.~\cite{Grier2}. Indeed, in that case, the values of $g$ and $k$
can be estimated as the inverses of mean times required for the
particle to diffuse a length $L/3-\sigma/2$ and $L/3$, respectively,
where $L$ is the intertrap distance and $\sigma$ is the width of an
optical trap.  Therefore, $g=2D/(L/3-\sigma/2)^2$ and $k=2D/(L/3)^2$,
$D$ being the diffusion coefficient of the colloidal sphere in the
fluid. Since $g>k$, the motion is from left to right if
$T^{(-1)}>T^{(+1)}$ and from right to left if $T^{(-1)}<T^{(+1)}$,
i.e., from each absorbing site in the longer-duration state toward the
nearest absorbing site in the short-lived state. The above expressions for $g$ and $k$ also show that the finite width of the optical traps plays an important role in the appearance of the steady directed motion.

Some limiting values of the expressions (\ref{velocity}) and
(\ref{effdiff}) can be easily interpreted. For instance, as suggested
by the previous simple considerations, $v$ and $D_{\mathrm{eff}}$ tend
to zero both as $\Gamma \rightarrow 0$ (absence of fluctuations
between the two lattice states) and as $\Delta \rho \rightarrow \pm 1$
(absence of fluctuations between the two lattice states in the
long-time limit). As $g\rightarrow 0$, $v$ and $D_{\mathrm{eff}}$ also
go to zero because, in this limit, the particle stays trapped between
two consecutive absorbing sites, since it is not allowed to jump to
them.  The expressions for $v$ and $D_{\mathrm{eff}}$ also tend to
zero both as $k\rightarrow 0$ and as $g\rightarrow \infty$. This is
due to the fact that, in these limits, the motion of the particle
consists of a series of forward and backward jumps from a
non-absorbing site to its nearest absorbing site, which arise as a
consequence of the fluctuations in the lattice state. Finally, the
limiting values $\lim_{\Gamma \rightarrow \infty} v=v_{\infty}$ and
$\lim_{\Gamma \rightarrow \infty}
D_{\mathrm{eff}}=D_{\mathrm{eff},\infty}$, which can be easily
calculated from Eqs.~(\ref{velocity}) and (\ref{effdiff}), also have a
simple interpretation. Using the results in Ref.~\cite{Derrida}, it is
not difficult to check that $v_{\infty}$ and $D_{\mathrm{eff},\infty}$
are the long-time velocity and the effective diffusion coefficient
corresponding to a random walk with the average transition rates
$L_n=\rho_{\mathrm{eq}}^{(+1)}L_n^{(+1)}+\rho_{\mathrm{eq}}^{(-1)}L_n^{(-1)}$
and
$R_n=\rho_{\mathrm{eq}}^{(+1)}R_n^{(+1)}+\rho_{\mathrm{eq}}^{(-1)}R_n^{(-1)}$.
In contrast with what happens in continuum models, the limiting value
$v_{\infty}$ does not vanish in general. This difference between
discrete and continuum models has been previously pointed out in the
literature (see, for example, Ref.~\cite{Ambaye}). From the
expressions for $v_{\infty}$ and $D_{\mathrm{eff},\infty}$, it can be
easily proved that $\lim_{g\rightarrow \infty} v_{\infty}=3 h \Delta
\rho(1-\Delta \rho^2) k/4$ and $\lim_{g\rightarrow
\infty}D_{\mathrm{eff},\infty} =9 h^2 (1-\Delta \rho^2) k/8$ which, in
general, are not equal to zero. This reflects the fact that the limits
$\Gamma \rightarrow \infty$ and $g \rightarrow \infty$ of
Eqs.~(\ref{velocity}) and (\ref{effdiff}) do not commute.

In order to simplify the graphical presentation of results, it is
convenient to introduce dimensionless quantities by defining
characteristic length and time scales. The characteristic length scale
$L$ is chosen to be the distance between two consecutive absorbing
sites, i.e., $L=3 h$.  The characteristic time scale $\tau$ is chosen
to be the mean time required for the particle to diffuse the length
$L$ in the absence of absorbing sites.  Taking into account that in
the absence of absorbing sites the effective diffusion coefficient is
$D_{\mathrm{eff,NA}}=h^2 k$, it follows that $\tau=L^2/(2
D_{\mathrm{eff,NA}})=9/(2 k)$. Then, the number of independent
dimensionless parameters is reduced to only three (the values of
$\tau\Gamma$, $\Delta \rho$, and $\tau g $), since the other two take
the fixed values $L^{-1}h=1/3$ and $\tau k=9/2$.

\begin{figure}
\includegraphics[width=8cm]{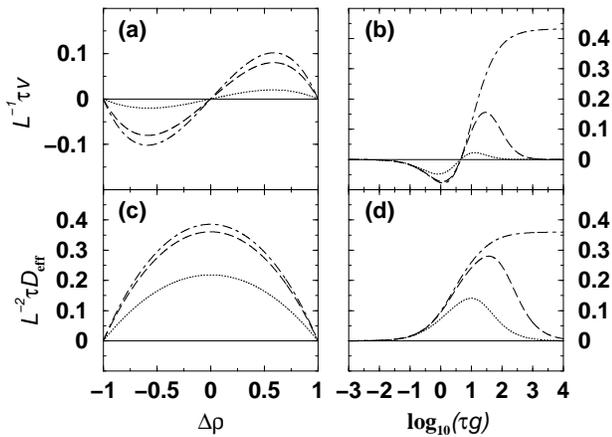}
\caption{\label{Fig3} Dependence of the dimensionless long-time
velocity $L^{-1}\tau v$ (top panels) and effective diffusion
coefficient $L^{-2}\tau D_{\mathrm{eff}}$ (bottom panels) on $\Delta
\rho$ (left panels) and $\log_{10}(\tau g)$ (right panels). The
dimensionless parameter values are: $L^{-1}h=1/3$, $\tau k=9/2$, $\tau
g=9$ (left panels), $\Delta \rho=0.6$ (right panels), $\tau \Gamma=0$
(solid lines), $\tau\Gamma =10$ (dotted lines), and $\tau\Gamma =100$
(long-dashed lines). The limiting values $ L^{-1}\tau v_{\infty}$ and
$L^{-2}\tau D_{\mathrm{eff,\infty}}$ are plotted with dash-dotted
lines.}
\end{figure}
In Fig.~\ref{Fig3}, we depict the dependence of the dimensionless
long-time velocity $L^{-1}\tau v$ (top panels) and effective diffusion
coefficient $L^{-2}\tau D_{\mathrm{eff}}$ (bottom panels) on $\Delta
\rho$ (left panels) and $\log_{10}(\tau g)$ (right panels). We have
taken $\tau g=9$ (left panels), $\Delta \rho=0.6$ (right panels), and
three values of $\tau \Gamma$, namely, $\tau \Gamma=0$ (solid lines),
$10$ (dotted lines), and $100$ (long-dashed lines).  With dash-dotted
lines we have plotted the limiting values $ L^{-1}\tau v_{\infty}$ and
$L^{-2}\tau D_{\mathrm{eff,\infty}}$.  Notice that both the absolute
value of $L^{-1}\tau v$ and $L^{-2}\tau D_{\mathrm{eff}}$ are
increasing functions of $\tau \Gamma$, except at those points at which
these functions vanish. This behavior is confirmed by a detailed study
of the signs of the derivatives $\partial v/\partial \Gamma$ and
$\partial D_{\mathrm{eff}}/\partial \Gamma$. The reversal of the
steady directed motion is clearly illustrated in panels (a) and
(b). These panels also reveal that, for fixed values of the remaining
parameters, there exist two optimal values of $\Delta \rho$ (opposite
to each other) and of $\tau g$ which maximize the absolute value of
the long-time velocity in each direction. In panel (b), one can also
see that the location of the second maximum is shifted towards larger
values of $\tau g$ as $\tau \Gamma$ is increased, and that this
maximum completely disappears in the limit $\Gamma \rightarrow
\infty$. Panel (c) shows that the dimensionless diffusion coefficient
as a function of $\Delta \rho$ is maximized for $\Delta \rho=0$, i.e.,
when the mean residence times in each lattice state are
equal. Analogously, there is also an optimal value of $\tau g$ which
maximizes the dimensionless diffusion coefficient as a function of
$\tau g$ [see panel (d)]. The location of this last maximum exhibits a
similar behavior with $\tau \Gamma$ as the second maximum in panel
(b).

\section{Conclusions}

\label{sect5}

In summary, we have proposed a simple random walk model which allows
the analytical study of the flux reversal in a spatially symmetric
fluctuating lattice. We have obtained explicit analytical expressions for the long-time velocity and the effective diffusion coefficient and made a study of them. 
In spite of its simplicity, by appropriately choosing the transition rates to an absorbing and a non-absorbing site,
our model is able to reproduce some important features of the experiment described in
Ref.~\cite{Grier2}. In particular, it provides the correct direction of the induced
flux and shows that this direction is closely related to the finite width of the optical traps. Furthermore, our model predicts the behavior of the effective diffusion coefficient that would be observed in the experiment reported in Ref.~\cite{Grier2}. Finally, it predicts that similar results to those contained
in \cite{Grier2} would be observed if the array of optical traps
switched randomly between its two possible configurations instead of periodically.

We have also presented an alternative interpretation of our model which 
mimics the motion of some molecular motors. In contrast to the one considered at the end of Ref.~\cite{Grier2}, we have proved that, in ours, the direction of the steady
motion depends not only on the relative durations of the two states,
but also on the sign of the difference between the transition rates to an absorbing and a non-absorbing site.

\begin{acknowledgments}
The author acknowledge the support of the Ministerio de Educaci\'on y 
Ciencia of Spain (FIS2005-02884) and the Junta de Andaluc\'{\i}a.
\end{acknowledgments}


\begin{thebibliography}{100}
\bibitem{Reimann} P. Reimann, Phys. Rep. {\bf 361}, 57 (2002).
\bibitem{Hanggi1} R. D. Astumian and P. H\"anggi, Phys. Today {\bf
  55}, 33 (2002).  
\bibitem{Bier} M. Bier and R. D. Astumian, Phys. Rev. Lett. {\bf 76},
  4277 (1996).
\bibitem{Prost} J. Prost, J. F. Chauwin, L. Peliti, and A. Ajdari,
Phys. Rev. Lett. {\bf 72}, 2652 (1994).
\bibitem{Chauwin} J. F. Chauwin, A. Ajdari, and J. Prost,
Europhys. Lett. {\bf 27}, 421 (1994); {\bf 32}, 373 (1995).
\bibitem{Rousselet} J. Rousselet, L. Salome, A. Ajdari, and J. Prost,
Nature (London) {\bf 370}, 446 (1994).
\bibitem{Faucheux} L. P. Faucheux, L. S. Bourdieu, P. D. Kaplan, and
A. J. Libchaber, Phys. Rev. Lett. {\bf 74}, 1504 (1995).
\bibitem{Reimann1} P. Reimann, Phys. Rev. Lett. {\bf 86}, 4992 (2001).
\bibitem{Doering} C. R. Doering, W. Horsthemke, and J. Riordan, Phys. Rev. Lett. {\bf 72}, 2984 (1994).
\bibitem{Elston} T. C. Elston and C. R. Doering, J. Stat. Phys. {\bf 83}, 359 (1996).
\bibitem{Chen} Y.-D. Chen, Phys. Rev. Lett. {\bf 79}, 3117 (1997).
\bibitem{Kanada} R. Kanada and K. Sasaki, J. Phys. Soc. Jpn. {\bf 68}, 3759 (1999).
\bibitem{Howard} J. Howard, {\it Mechanical of Motor Proteins and the
Cytoskeleton} (Sinauer Associates, Sunderland, 2001).
\bibitem{Grier1} S.-H. Lee, K. Ladavac, M. Polin, and D. G. Grier,
Phys. Rev. Lett. {\bf 94}, 110601 (2005).
\bibitem{Grier2} S.-H. Lee and D. G. Grier, Phys. Rev. E {\bf 71},
060102(R) (2005).
\bibitem{Kolomeisky} A. B. Kolomeisky and B. Widom,
J. Stat. Phys. {\bf 93}, 633 (1998).
\bibitem{Freund} J. A. Freund and L. Schimansky-Geier, Phys. Rev. E
{\bf 60}, 1304 (1999).
\bibitem{Ambaye} H. Ambaye and K. W. Kehr. Physica A {\bf 267}, 111
(1999).
\bibitem{Makhnovskii} Y. A. Makhnovskii, V. M. Rozenbaum, D.-Y. Yang, 
S. H. Lin, and T. Y. Tsong, Phys. Rev. E {\bf 69}, 021102 (2004).
\bibitem{prove} Indeed, if $N=2$, from the periodicity and the
symmetry property it follows that $R_1^{(\alpha)}=L_1^{(\alpha)}$ and
$R_2^{(\alpha)}=L_2^{(\alpha)}$. Consequently, from Eq.~(\ref{vtdef})
one has that $v=0$. Analogously, if $N=1$, $v=0$ trivially.
\bibitem{Derrida} B. Derrida, J. Stat. Phys. {\bf 31}, 433 (1983).



\end{thebibliography}
\end{document}